\documentclass[reprint,amsmath,amssymb,aps,showpacs]{revtex4-2}  
\pdfoutput=1

\usepackage{graphicx}
\usepackage{dcolumn}
\usepackage{bm}  
\usepackage{xcolor}
  
\begin{document}

\title{Zoology of collective patterns \\ modulated by non-reciprocal, long-range  interactions}


\author{Edgardo Brigatti}  
\affiliation{Instituto de F\'{\i}sica, Universidade Federal do Rio de Janeiro, 
Av. Athos da Silveira Ramos, 149,
Cidade Universit\'aria, 21941-972, Rio de Janeiro, RJ, Brazil}
\email{edgardo@if.ufrj.br}
\author{Fernando Peruani}
\affiliation{Laboratoire de Physique Th\'eorique et Mod\'elisation, UMR 8089, CY Cergy Paris Universit\'e, 95302 Cergy-Pontoise, France.}
\email{fernando.peruani@cyu.fr}


\begin{abstract}

We investigate active particles that exhibit long-range interactions only restricted by a field of view, which is characterized by an angle $\beta$.  
We show that constraining attractive interactions 
to a field of view leads to the emergence of a complex pattern that exhibits -- depending on the value of $\beta$ and initial conditions -- significantly different topologies and transport properties. 
We find, in two dimensions, a nematic closed filament in the form of a ring that moves as a chiral active particle, a closed polar filament with one singular topological point that exhibits net polar order and moves ballistically, a structure with two singular topological points that rotates, or an open polar filament that behaves as a persistent random walk. 
Furthermore, we investigate the process that transforms one structure into another by slowly varying $\beta$ and observe that the process is non-reversible and presents strong hysteresis.  
Finally, we find that in three dimensions similar patterns also emerge.  
The analysis sheds light on the physics of single-species active particles with long-range, non-reciprocal interactions in two and three dimensions, characterized by the absence of gas phases, and provides evidence that in these systems, topological and transport properties are closely related. 

\end{abstract}

\maketitle

Collective motion patterns, such as flocking or milling, observed in biological systems -- including birds, fish, and sheep~\cite{vicsek2012,marchetti2013,ballerini2008,gautrais2012,ginelli2015,toulet2015} -- or in artificial active systems~\cite{grossman2008,deseigne2010,weber2013,dauchot2015}, are almost always explained by invoking the apparent necessity of an underlying velocity alignment mechanism that mediates interactions among actively moving entities. 
Velocity alignment is a central concept in polar active fluids --  e.g. in the Vicsek model~\cite{vicsek1995} -- as well as in active nematics. This alignment mechanism is inspired by the XY model and assumes that velocities interact as spins. Such an analogy makes velocity alignment particularly theoretically appealing~\cite{vicsek2012,marchetti2013}.  
Despite this connection to the XY model, the intrinsic non-equilibrium nature of active systems leads to fundamental differences, such as the emergence of long-range orientational order in two-dimensions~\cite{vicsek1995,toner1995,toner1998} or 
the presence of anomalous density fluctuations~\cite{ramaswamy2003,ramaswamy2010}. 

\begin{figure}
\begin{center}
\resizebox{\columnwidth}{!} {\includegraphics{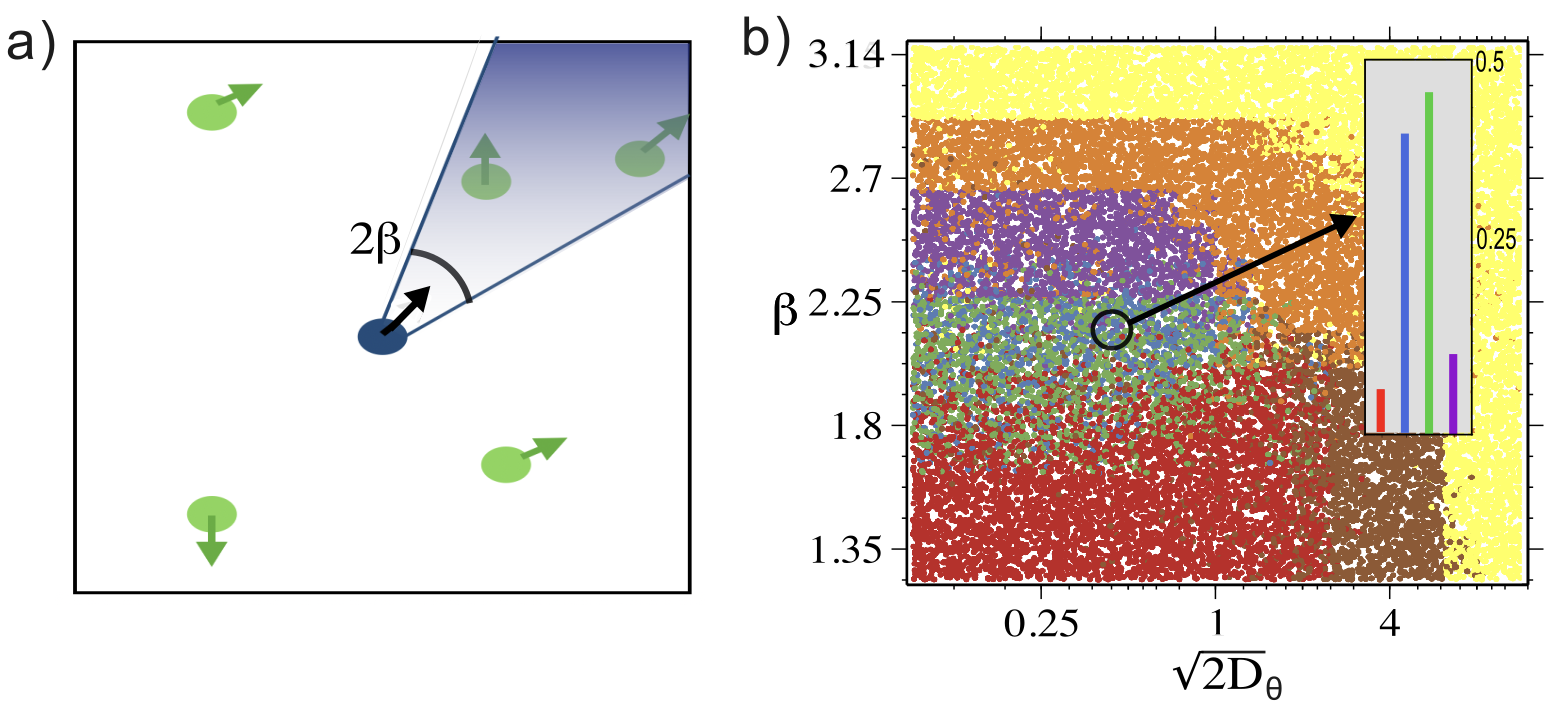}}
\caption{
(Color online) (a) Scheme of the model: the vision cone of the black particle is the blue region defined by the angle $2\beta$.
(b) Phase diagram varying the angle $\beta$ and the noise intensity $D$. 
The inset displays the probability -- at the position indicated by the circle and starting from random initial conditions -- of observing various  patterns (that are color coded). Color code: worms (red), 2-twist (green), 3-twist (blue), ring (violet),  planet (orange), amoeba (brown) and cloud (yellow); for more details see \cite{SuppInf}. 
 }
\label{fig:1}
\end{center}
\end{figure}

\begin{figure*}[!]
\begin{center}
\resizebox{\textwidth}{!} {\includegraphics{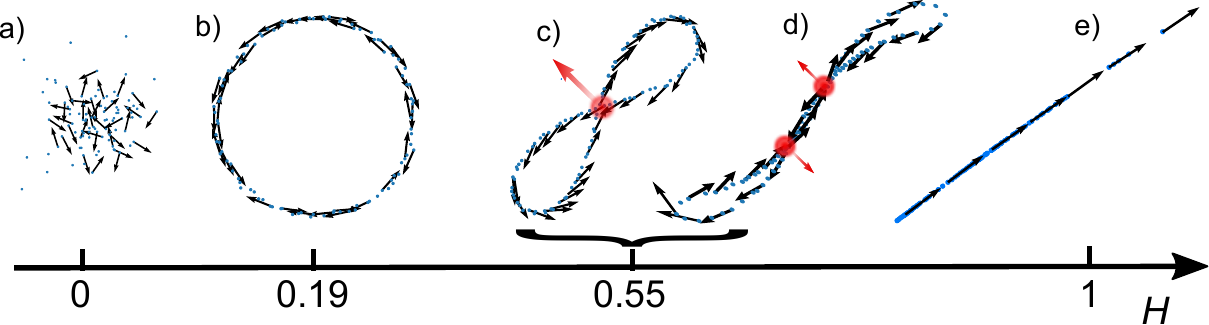}}
\caption{
Zoology of collective moving patterns as the non-reciprocity index ${H}$ is increased:
a) a cloud ($\beta$=$\pi$), 
b) a ring ($\beta$=2.45), 
c) a 2-twist ($\beta$=2.06), 
 d)a 3-twist ($\beta$=1.93),
 and e) a worm ($\beta$=1.54).For all patterns the noise amplitude is $\sqrt{2 D}=0.12$. 
The red points -- in c) and d) -- indicate  singular topological points 
and the red arrows the resulting local polar order at those points. For more information and illustrative videos of each emergent collective pattern, see \cite{SuppInf}. 
 }
\label{fig:2}
\end{center}
\end{figure*}

Interestingly, collective motion patterns, such as flocking or milling, can emerge even in the absence of velocity alignment~\cite{romanczuk2009,strombom2011,moussaid2011,pearce2014,huepe2013,huepe2015, barberis2016,grossmann2013,soto2014}.
Particularly relevant for applications to animal systems~\cite{calvao2014,barberis2016,calvao2019}, pedestrian models~\cite{moussaid2011}, and realistic vision-based models~\cite{bastien2020model,castro2024modeling} 
is the notion that collective organization can emerge from simple rules based on the position, and not the velocity, of the neighbors. 
Moreover,  a simple attraction rule has been shown to lead to the emergence of complex collective patterns, beyond standard aggregation, if and only if the interactions among identical particles break the action-reaction symmetry, for example, by restricting perception via a field of view~\cite{barberis2016}.

On the other hand, action-reaction symmetry breaking has been shown to play a major role in (scalar) ``active mixtures" -- i.e. active systems with two or more particle types -- where non-reciprocal interactions between particle types lead to non-equilibrium patterns. 
Examples range from the emergence of traveling patterns~\cite{you2020nonreciprocity} and non-equilibrium self-assembly~\cite{soto2014self} to  phase coexistence~\cite{dinelli2023non} and chaotic dynamics of bands for quorum-sensing interactions~\cite{duan2023dynamical}. 
A major theoretical interest in (scalar) active mixtures has been that the effects of non-reciprocal interactions can be understood at the hydrodynamic level by constructing Cahn-Hilliard-type models for non-reciprocal interactions~\cite{saha2020scalar,frohoff2023nonreciprocal,you2020nonreciprocity}.
However, single-species, non-reciprocal active systems -- despite being ubiquitous in real-world systems such as in sheep~\cite{gomez2022intermittent}, birds~\cite{nagy2010hierarchical},  starfish embryos~\cite{tan2022odd} or robots~\cite{soria2019influence} -- remain, comparatively,  unexplored.
Exceptions include: observations of long-range order in two dimensions for XY spins with vision cones~\cite{loos2023long}, recently found to be metastable~\cite{dadhichi2020nonmutual, popli2025don, dopierala2025inescapable}, 
order in the presence of velocity alignment with fore-aft asymmetry~\cite{chen2017fore}, 
collective patters of attractive active particles with vision cone~\cite{barberis2016}, including adaptive attraction ~\cite{li2019role} and steric repulsion~\cite{negi2022emergent,negi2024collective} and aggregation patterns of torque-free particles~\cite{lavergne2019group}. 
Finally, hydrodynamic equations for these systems are limited to those derived for attractive active particles with a vision cone~\cite{barberis2016,peruani2017hydrodynamic} and those with non-reciprocal velocity alignment~\cite{huang2024active}. 

Here, for the first time, we investigate the emergence of collective patterns in single-species, active systems with long-range (attractive) interactions restricted solely by a  field of view [Fig.~\ref{fig:1}(a)]. 
Note that previous models including a field of view~\cite{barberis2016,li2019role,negi2022emergent,negi2024collective,lavergne2019group} considered a finite horizon beyond which interactions do not occur and thus effectively restricted interactions to be of short range (with the exception of  ~\cite{calvao2014,calvao2019}).  

Importantly, long-range, attractive interactions lead to two fundamental properties -- not present in systems with short-range interactions -- that strongly impact the emergent dynamics. First, despite the restriction imposed by the vision cone, a single particle never loses contact with the other particles. As a result, 
the system remains cohesive even in the limit of vanishing density and for all noise intensities. 
In other words, a gas phase  of freely expanding particles in open space does not exist.  In sharp contrast, 
active particles with short-range interactions always exhibit a gas phase. 
Second, long-range interactions allow particles to form a single stable complex pattern whose size and topology result from the system dynamics and thus are not determined/constrained by a characteristic interaction length. 

In short, our minimal framework -- that is not intended as a realistic vision-based model -- provides insight into the complex dynamics that emerges in collectives with long-range, non-reciprocal attraction. The results obtained may help to understand collective behavior in animal groups, robot swarms, and chemotactic active colloid systems.




{\it Model--} We consider a system of $N$ active particles that move at 
constant speed. 
In any spatial dimension, the equation of motion of the $i$-th particle, with position ${\bf x}_i$,  is given by: 
\begin{equation}
\label{eq:3d}
\ddot{\mathbf{x}}_i 
= \hat{O}(\dot{\mathbf{x}}_i)\,\mathbf{F}_i+  \hat{O}(\dot{\mathbf{x}}_i)\circ\mathbf{N}_i \, .
 \end{equation}
$\mathbf{F}_i = \frac{\gamma}{n_i} 
   \sum_{j \in \Omega_i} 
   \frac{\mathbf{x}_j - \mathbf{x}_i}{\lVert \mathbf{x}_j - \mathbf{x}_i \rVert}$ 
is the interaction force: particle $i$ is attracted to all particles $j$ within its field of view. 
The term $\Omega_i$ denotes the set of all particles present in the field of view of 
$i$ at time $t$ [see Fig. \ref{fig:1}(a)], and $n_i$ is its cardinality (the number of neighbors of $i$). The parameter $\gamma$ is the relaxation constant. 
The projector operator $\hat{O}(\dot{\mathbf{x}}_i)$  applied to a vector $\mathbf{A}$ is defined by $\hat{O}(\dot{\mathbf{x}}_i)\, {\mathbf{A}} \equiv -\frac{1}{v_0} \dot{\mathbf{x}}_i \times (\dot{\mathbf{x}}_i\times \mathbf{A})$. The operator $\hat{O}$ 
ensures that $||\dot{\mathbf{x}}_i||$ remains constant by eliminating the component of $\mathbf{A}$ parallel to $\dot{\mathbf{x}}_i$ and leaving only the component perpendicular to it. The term $\mathbf{N}_i$ denotes a vector noise with components $N_{i\alpha}(t)$ ($\alpha \in \{x,y,z\}$) such that $\langle N_{i\alpha}(t)N_{i\beta}(t') \rangle = 2D \delta_{\alpha\,\beta} \delta(t-t')$. Since the projector operator $\hat{O}$ depends on $\dot{\mathbf{x}}_i$, the noise is interpreted in the Stratonovich sense. 

In the following, we focus on two dimensions. The behavior of the system in three dimensions is discussed later. The active model with attractive, non-reciprocal interactions given in Eq.~(\ref{eq:3d}) reduces in 2D to:
\begin{eqnarray}
\label{eq:x}
\dot{{\bf x}}_i &=& v_0\hat{\bf{e}}[\theta_i]  \, , \\
\label{eq:theta}
\dot{\theta}_i &=& \frac{\gamma}{n_i}\sum_{j\in \Omega_i} \sin{(\alpha_{ij}-\theta_i)}+\sqrt{2D}\xi_i(t) \, ,
\end{eqnarray} 
where, in Eq.~(\ref{eq:x}), $\theta_i$ encodes the velocity direction using $\hat{\bf{e}}[\cdot] \equiv \cos(\cdot)\hat{x} + \sin(\cdot)\hat{y}$, while in  Eq.~(\ref{eq:theta}), $\alpha_{ij}$ is the polar angle associated with the vector  $({\bf x}_j- {\bf x}_i)/|| {\bf x}_j- {\bf x}_i|| = \hat{\bf{e}}[\alpha_{ij}]$, 
and $\xi_i(t)$ is a white noise with  $\langle \xi_i(t) \rangle = 0$ and  $\langle \xi_i(t)  \xi_j(t') \rangle = \delta_{ij}\delta(t-t')$.  
%
The set $\Omega_i$ is defined by the following condition: any particle $j$ such that  $\hat{\bf{e}}[\alpha_{ij}]\cdot\hat{\bf{e}}[\theta_i]>\cos(\beta)$ 
belongs to the field of view of the particle $i$, which does not involve any restriction on the distance between $i$ and $j$, as occurs in~\cite{barberis2016}. Note that $\beta$ controls the size of the vision cone.   
Since the equations can be adimensionalized by choosing an adequate length and time scale,  we fix $v_0=1$ and $\gamma=5$ without loss of generality.

{\it General remarks--} 
The force $\mathbf{F}_i$ in Eq.~(\ref{eq:3d}) can be expressed as $\mathbf{F}_i = - \frac{\partial U_i}{\partial \mathbf{x}_i}$ with $U_i = \frac{\gamma}{n_i}\sum_{i=1}^{n_i} || \mathbf{x}_j - \mathbf{x}_i ||$. 
One of the fixed points of  Eq.~(\ref{eq:3d}) is reached at the minimum of this potential and corresponds to a vanishing interparticle distance. However, note that the constant speed constraint, among other features of the system dynamics, prevents particles from collapsing into a single point.  

For $\beta=\pi$, it can be shown that particles are attracted to a (weighted) center of mass.  
In this reciprocal limit, i.e. $\beta=\pi$, the spatial distribution of particles stays asymptotically confined within a radius $0<\epsilon<\infty$, with $\epsilon=\epsilon(D,\gamma,v_0)$. 
Consequently, there is no gas phase 
with particles 
freely expanding in open space. In contrast,  a system of active particles with short-range interactions possesses such a gas phase. 

This result also holds 
for $0<\beta<\pi$, i.e. when particles display a blind angle, and thus interactions are non-reciprocal. 
Specifically, we argue that freely diffusive particles cannot exist for $0<\beta \leq \pi$ and $D>0$:  a particle taken away from a high-density swarm -- independently of the morphology of the collective --  tries to move back to it. 
%
%
Assume that a particle is in position $\mathbf{x}_A$ and a swarm is localized at position $\mathbf{x}_B$. For simplicity, let us consider that $\mathbf{x}_B$ is constant. 
%
If 
$\mathbf{x}_B$ is not in the field of view of $A$, then $\theta_A$ will follow a purely diffusive dynamics. With certainty, in finite time $\tau$ -- with $\langle \tau \rangle \leq (\pi-\beta)/D$) in 2D -- $B$ will be in the field of view of $A$ and will start to move towards $B$.  
In summary, $A$ performs a (persistent) random walk when $B$ is not in its field of view, while it moves ballistically towards $B$ when $B$ is detected. 
For any 
positive value of $D$, the dynamics is such that it prevents $A$ to become independent of $B$ and thus inhibits the free expansions of the  spatial distribution of particles.  
Thus, we can ensure the absence of a gas phase for our system of active particles with long-range interactions in 2 and 3 dimensions.

{\it Emergent self-organized patterns--} 
As commented above, starting from a random initial spatial distribution, particles self-organize into a single pattern/structure.  
Interestingly, the emergent structure can display various topologies [see Fig.\ref{fig:2}].  
As in the Game of Life~\cite{adamatzky2010game}, the initial configuration of the particles plays a crucial role in the collective pattern that we  observe [see Fig.\ref{fig:1}(b)].  
Moreover, we find that for a fixed parameter set, 
several attraction basins  -- one for each collective pattern -- coexist. 
Importantly, the transport properties of the center of mass (CM) of these patterns differ from one structure to another: the CM can diffuse, move ballistically (for a very long characteristic time), or the structure rotates.  
Interestingly, there is a clear connection between the topology of the 
pattern and its transport properties. 

A fundamental aspect of these complex structures is that they can only emerge in the absence of the action-reaction symmetry, which is broken by the field of view~\cite{commentAR}. 
The level of non-reciprocity is then a key feature 
and it differs from structure to structure.  
To characterize how non-reciprocal interactions are, we introduce the non-reciprocity index $H$, defined as: 
\begin{equation}
H(t)=\frac{1}{K}\sum_{i<j} |A_{i,j}(t)-A_{j,i}(t)|,
\end{equation}
where, at time $t$, the adjacency matrix element is defined as $A_{i,j}=1$ if particle $j$ is in the vision cone of $i$ and $0$ otherwise.  
$K$ is a normalization constant ($K=N\,(N-1)/2$) which ensures that $H(t) \in [0,1]$. 
When all interactions are reciprocal, e.g. for $\beta=\pi$, $A_{i,j}(t)=A_{j,i}(t)$ for all $(i,j)$, and thus $H(t)=0$. 
On the other hand, for non-reciprocal interactions $|A_{i,j}(t)-A_{j,i}(t)|=1$. If this applies to all pairs $(i,j)$, then $H(t)=1$. 
We characterize emergent self-organized structures by computing $H=\langle H(t) \rangle_t$, with $\langle \cdots \rangle_t$ denoting the temporal average. 
Notably, the value of $H$ depends not only on the vision cone angle $\beta$, but also on the topology of the structure. 
We further analyze the interaction network $A_{i,j}(t)$  
by measuring the temporal evolution of the 1-norm between the adjacency matrix at time $t_0$ and at time $t>t_0$: $d_1=\sum_{i}\sum_{j} | A_{i,j}(t_0)-A_{i,j}(t) |$. 
 
Furthermore, we study the phase portrait $[\theta_i, \dot{\theta}_i]$ and the transport properties of the collective pattern. To do that, we compute the temporal evolution of the CM, defined as ${\bf x}_{CM}(t) = \sum_i {\bf x}_i(t)/N$, its average squared $\delta^2(t) = \langle \left({\bf x}_{CM}(t_0+t) -   {\bf x}_{CM}(t_0) \right)^2\rangle_{t_0}$, the polarization $P(t) = |{\bf P}(t)| = |\sum_i {\bf{e}}(\theta_i(t))/N|$, and its correlation $C(t) = \langle {\bf P}(t_0+t) \cdot {\bf P}(t_0) \rangle_{t_0}$. 
 Note that $\dot{{\bf x}}_{CM} = v_0 {\bf P}(t)$ and thus ${\bf x}_{CM}(t) = v_0  \int {\bf P}(t') dt'$, implying that $\delta^2(t)  = 2 \int_0^{t} dt_1 \int_0^{t_1} dt_2 C(t_1-t_2)$.   

\begin{figure}
\begin{center}
\resizebox{\columnwidth}{!} {\includegraphics{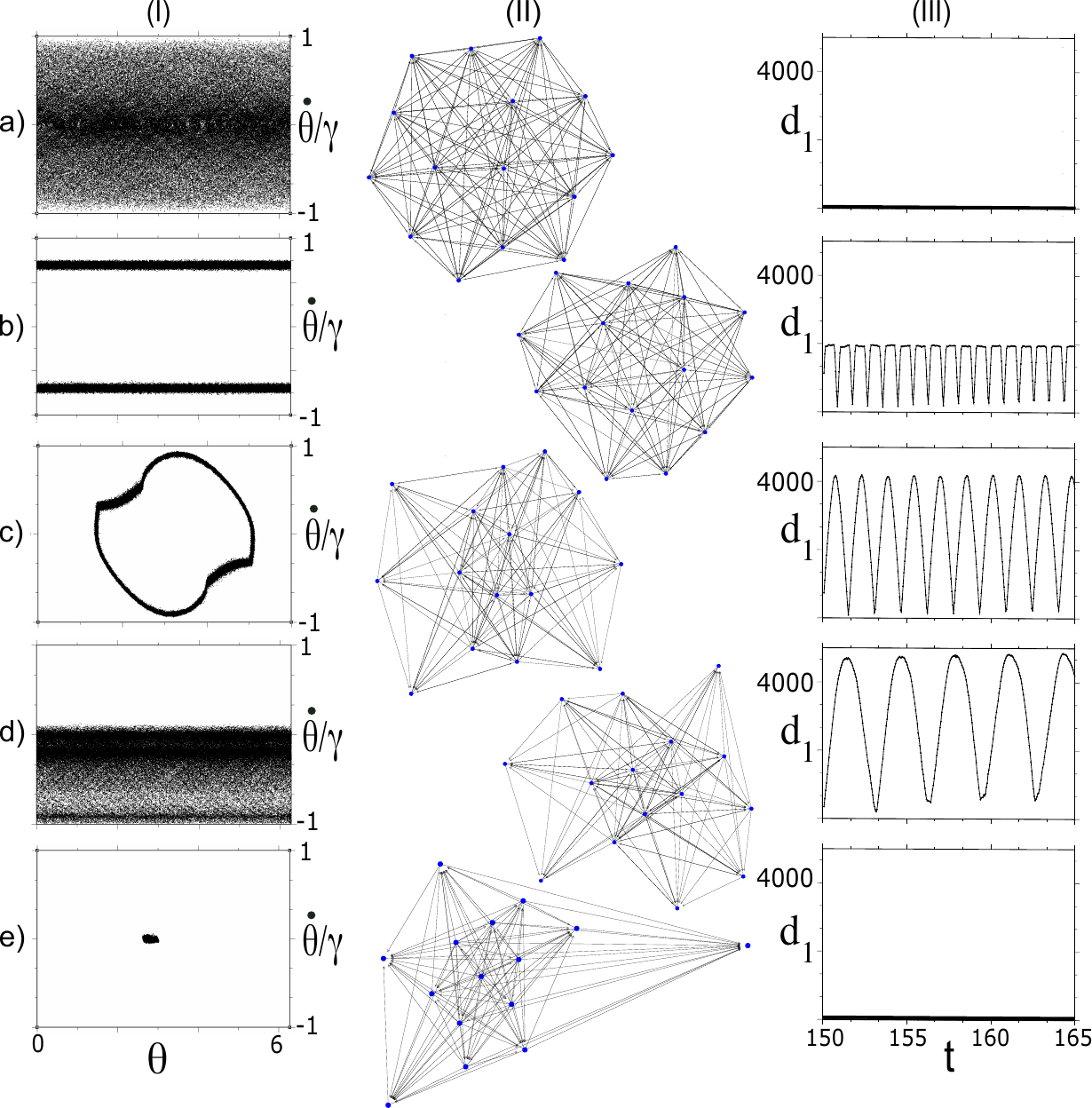}}
\caption{
Columns correspond to the phase portrait (I), the interaction network (II), and the interaction network dynamics (III) of the collective pattern. From top to bottom: clouds (a), rings (b), 2-twists (c), 3-twists (d) and worms (e). Vision angle values $\beta$ are given in Fig.~\ref{fig:2}. Noise amplitude $\sqrt{2D}=0.12$. For more details, see~\cite{SuppInf}.
 }
\label{fig:3}
\end{center}
\end{figure}

In the following, we describe the most representative structures and their main properties for increasing non-reciprocity index $H$; see \cite{SuppInf} for illustrative videos of each 
collective pattern and more technical information. 

\begin{figure}
\begin{center}
\resizebox{\columnwidth}{!} {\includegraphics{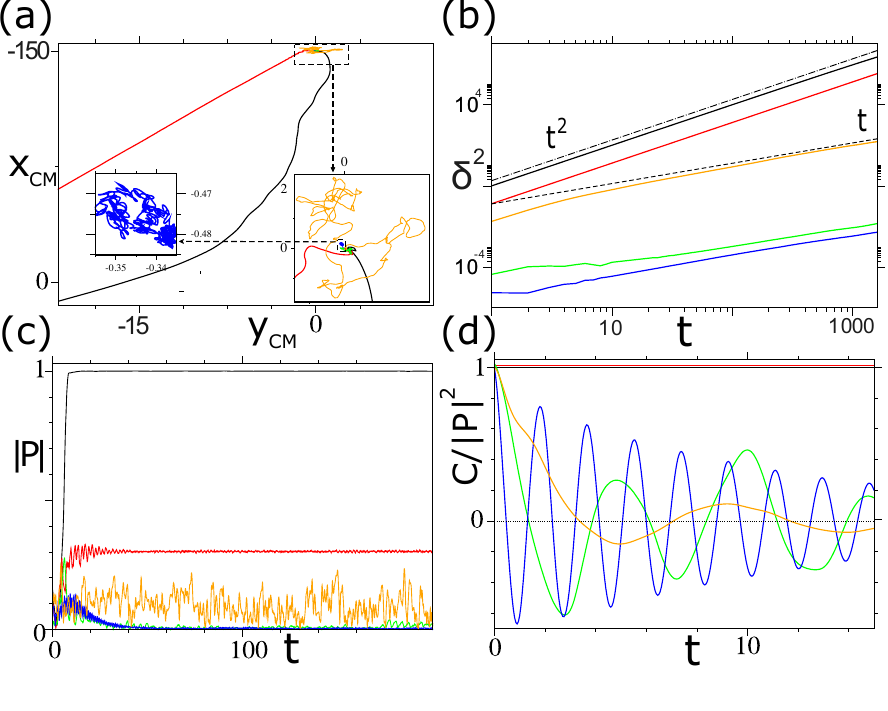}}
\caption{
Transport properties: trajectories of the CM for different collective patterns (a), 
mean quadratic distance of CM vs time (b), modulus of the polarization vs time (c) and autocorrelation of the direction of the polarization (d).
Color code: worms (black lines),  2-twist (red),  3-twist (green),  ring (blue) and cloud (orange). Vision angle values $\beta$ are given in Fig.~\ref{fig:2}. Noise amplitude $\sqrt{2D}=0.12$. 
 }
\label{fig:4}
\end{center}
\end{figure}

{\it Cloud.--} For $\beta=\pi$ interactions are reciprocal, and thus $H = 0$.  
The particles orbit around the CM and the pattern appears as a roundish cloud [Fig.\ref{fig:2}]. 
The resulting interaction network is a fully connected static network [Fig.\ref{fig:3}, row a)]. The phase portrait $[\theta_i, \dot{\theta}_i]$ is homogeneously covered [Fig.\ref{fig:3}, row a)]. Particles move along elongated 8-shaped trajectories around the CM \cite{SuppInf}. This cloud of particles has a vanishing polarization. Fluctuations, due to $D>0$, lead asymptotically to the diffusive behavior of the CM [see Fig.\ref{fig:4}].  

{\it Ring.--} When $\beta$ is slightly smaller than $\pi$ and such that $H \sim 0.19$, particles self-organize into ring patterns [Fig.\ref{fig:2}]. 
The interaction network remains almost fully connected and exhibits a clear 
oscillatory dynamics [Fig.\ref{fig:3}, row b)]. 
Rings consist of 50\% of the particles, homogeneously distributed along the ring, rotating clockwise, while the other 50\% rotating counter-clockwise. 
This implies that locally, in a small segment of the ring, half of the particles move in one direction and the other half in the opposite direction. 
The local polar order vanishes, and thus there is zero global polarization. The ring is a closed nematic filament.  
This is evident by looking at the phase portrait 
that shows that $\dot{\theta}_i$ is $0.7\gamma$ or $-0.7\gamma$, while $\theta_i$ is homogeneously distributed over $[0, 2\pi)$; [Fig.\ref{fig:3}, row b)]. 
The period of a particle turning around the ring is $\frac{2\pi}{0.7\gamma}$, and the period exhibited by the interaction network is, as expected, half of this value. 
Since the radius $R$ obeys $v_0=\dot{\theta}_i\,R$, then $R = \frac{v_0}{0.7\gamma}$. 
The presence of a blind angle implies that each particle does not move towards the CM, but to a point slightly displaced away from the CM. 
Furthermore, we observe that the CM rotates. 
This rotation is noisy because of the angular fluctuations experienced by each particle. 
We find that the behavior of the CM is consistent with a chiral random particle model such that $\dot{{\bf x}}_{CM}= v_0 {\bf P}= v_{R} \hat{{\bf e}}({\theta_{R})}$ and $\dot{\theta}_{CM} = \Omega_{R} + \sqrt{2D_{R}} \eta(t)$, with $v_{R}$, $\Omega_{R}$, and $D_{R}$ constant and $\eta(t)$ a white noise [Fig.\ref{fig:4}]. 
Asymptotically, the behavior of CM is diffusive with diffusivity~\cite{otte2021statistics} $D = \frac{v_{R}^2 D_{R}}{2(\Omega_R^2 + D^2_{R}) }$, which is much smaller than the diffusivity of the cloud.  

{\it 2-Twist.--} 
Further decreasing the value of $\beta$, several stable complex patterns can emerge depending on the initial condition. 
One of them is an 8-shaped pattern, which we call the 2-Twist pattern [Fig.\ref{fig:2}], commonly observed at non-reciprocity index $H \sim 0.55$ and displaying a relatively complex interaction network with intermediate connectivity values [Fig.\ref{fig:3}, row c)].
The 2-Twist pattern consists of particles moving along this 8-shaped orbit always in the same direction, implying that the particles exhibit local polar order. 
The time a particle takes to move along this orbit sets the period observed in the periodicity of the interaction network. 
The phase portrait 
corresponds to a non-trivial closed orbit that reflects a complex oscillatory behavior of $\theta_i$ as the particle moves along the structure. 

The remarkable feature of this pattern is the presence of a singular topological point where the derivative of the polarization along the structure exhibits a discontinuity [see Fig.\ref{fig:2}]. 
This point corresponds to the crossing of two segments of the polarized filament.  
Note that closed polar filaments with no crossing cannot display (global) polar order. 
However, if the self-organized structure has a crossing, i.e. a singular topological point, the structure can exhibit non-zero polar order. 
The polar order displayed by the structure [see Fig.\ref{fig:4}] is given by the polar order at the singular topological point.
As indicated above, $\dot{{\bf x}}_{CM} = v_0 {\bf P}$. 
The high temporal correlation value displayed by polar order ${\bf P}$ 
implies that the CM moves ballistically for a long characteristic time [Fig.\ref{fig:4}]. 
Arguably, angular fluctuations should render CM motion asymptotically diffusive, but the persistence time seems to be extremely large, to the point that we failed to observe it in simulations.  

{\it 3-Twist.--} 
For parameter values where the 2-Twist pattern is observed, we can also find
another structure that displays not one, but two singular topological points [Fig.\ref{fig:2}]. We call this structure 3-Twist. 
The non-reciprocity index $H \sim 0.55$ is close to the one measured in 2-Twist patterns. The interaction network presents a connectivity similar to the 2-Twist case [Fig.\ref{fig:3}, row d)] but oscillates with a longer period (these patterns are longer and the particles take longer time to move around them).  
Only half of the phase portrait is occupied by a structured spatial coverage determined by the wave-like dynamics of the particles trajectories (details in \cite{SuppInf}).
Importantly, the 3-Twist pattern corresponds to a closed polar chain of particles that displays two crossings, i.e. topological singular points.  
The magnitude of polarization at these two points ($|{\bf P}_{sp}|$) is the same and, given the topology of the structure, the local polar order, at each 
point, is opposite to the other. This implies 
that polar order is nearly zero and that the velocity of the CM vanishes.  
On the other hand, these topological singular points are separated by a distance $\ell$ and 
since each of them moves with speed  $v_0|{\bf P}_{sp}|$, the structure rotates around its CM. 
The angular velocity of this rotation is proportional to $v_0|{\bf P}_{sp}|/\ell$. 
%
While particles move along the 3-Twist structure, the oscillations of $\theta_i$, combined with the rotation of the structure itself \cite{SuppInf}, lead to a phase portrait pattern that fills half of the plane [Fig.\ref{fig:3}, row d)]. 
Finally, while the average velocity of the CM is $0$, angular fluctuations in the equations of motion lead asymptotically to diffusive behavior of the CM [Fig.\ref{fig:4}]. 

\begin{figure}
\begin{center}
\resizebox{\columnwidth}{!} {\includegraphics{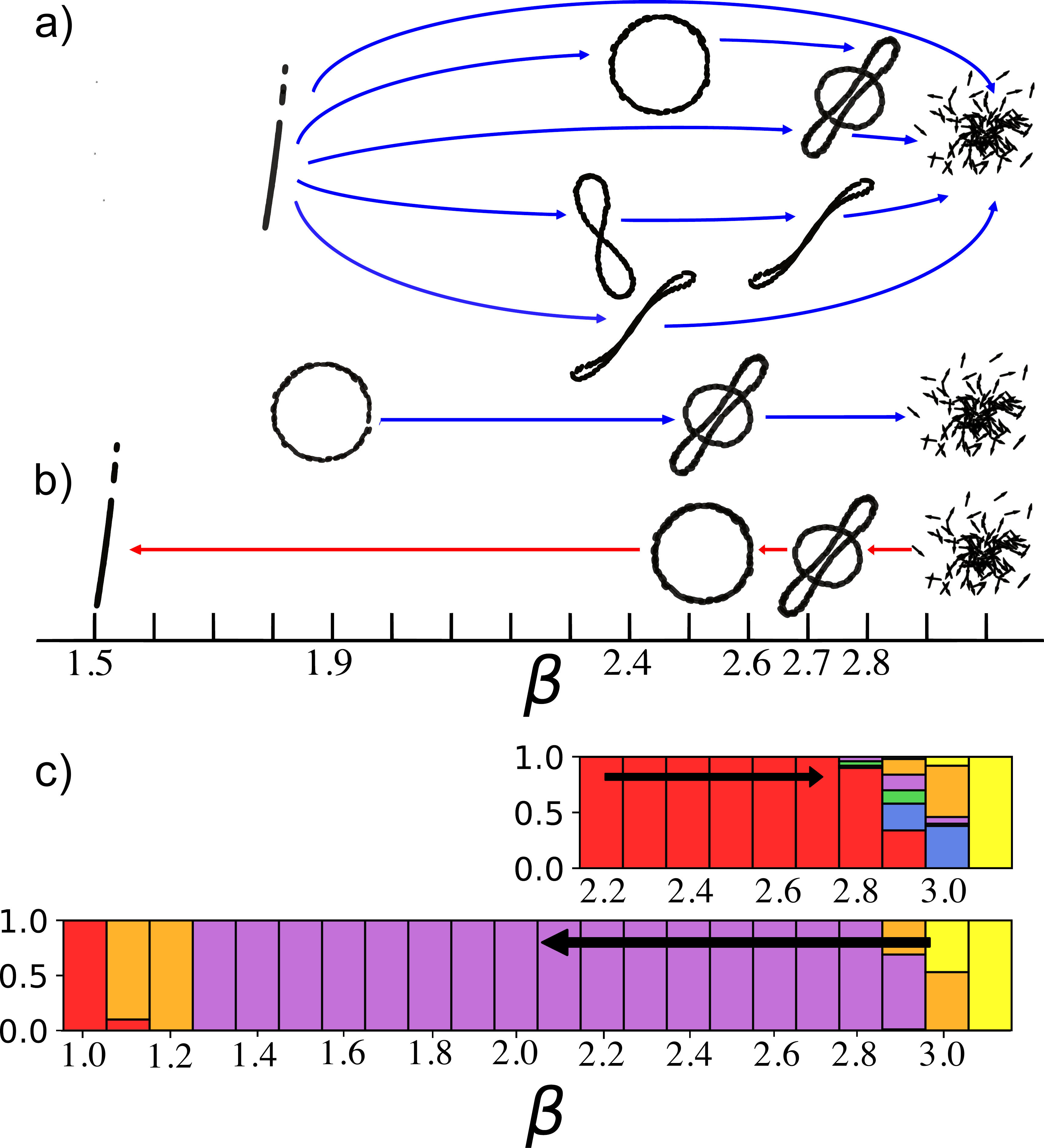}}
\caption{
Evolution of a collective motion pattern as the field of view $\beta$ is varied. (a) Above: transition from a worm to a cloud pattern. As $\beta$ is varied, different patterns can emerge. There is no unique sequence of patterns. Below: transition from a ring to a cloud
(b) The sequence of patterns observed varying $\beta_1 \to \beta_2$ is different from $\beta_2 \to \beta_1$, indicating  hysteresis effects. 
(c) Probabilities of finding the considered collective patterns tuning the $\beta$ value for the cases displayed in (a) -- from a worm to a cloud, increasing $\beta$ -- and in (b) -- decreasing $\beta$--. Same color code of Fig.1. For more details, see~\cite{SuppInf}.    
 }
\label{fig:5}
\end{center}
\end{figure}

{\it Worm.--} 
At lower values of $\beta$, other distinct self-organized patterns with non-reciprocal index $H \sim 1$ emerge: open polar filaments, which correspond to files of active particles following each other [Fig.\ref{fig:2}]. We call this collective pattern worm. 
The resulting interaction network is static, highly non-reciprocal and hierarchical [Fig.\ref{fig:3}, row e)]. The particles interact with those
in front of them in the filament. Thus, the particle in the front does not interact with anybody, the second particle in the filament with the particle at the front, the third particle with the second and first particle, 
and so on. 
In this structure  $\langle \dot{\theta}_i \rangle =0$ and, for all $i$, $\theta_i(t) \sim \theta_*(t)$, where $\theta_*(t)$ is the angular variable of the particle in front of the worm at time $t$. Thus, the phase portrait is approximately reduced to a fixed point. Worms display high polar order with $|{\bf P}|\sim 1$. 
At a given time $t$, we can roughly assume that ${\bf P}(t) \parallel \hat{\bf{e}}[\theta_*(t)]$.
This means that the stochastic dynamics of $\theta_*$ is followed by all particles and therefore the CM moves ballistically during a characteristic persistent time $D^{-1}$ [Fig.\ref{fig:4}].  
Only on much longer timescales can motion be recognized as diffusive, as is expected for an isolated particle. 

\begin{figure*}[!]
\begin{center}
\resizebox{\textwidth}{!} {\includegraphics{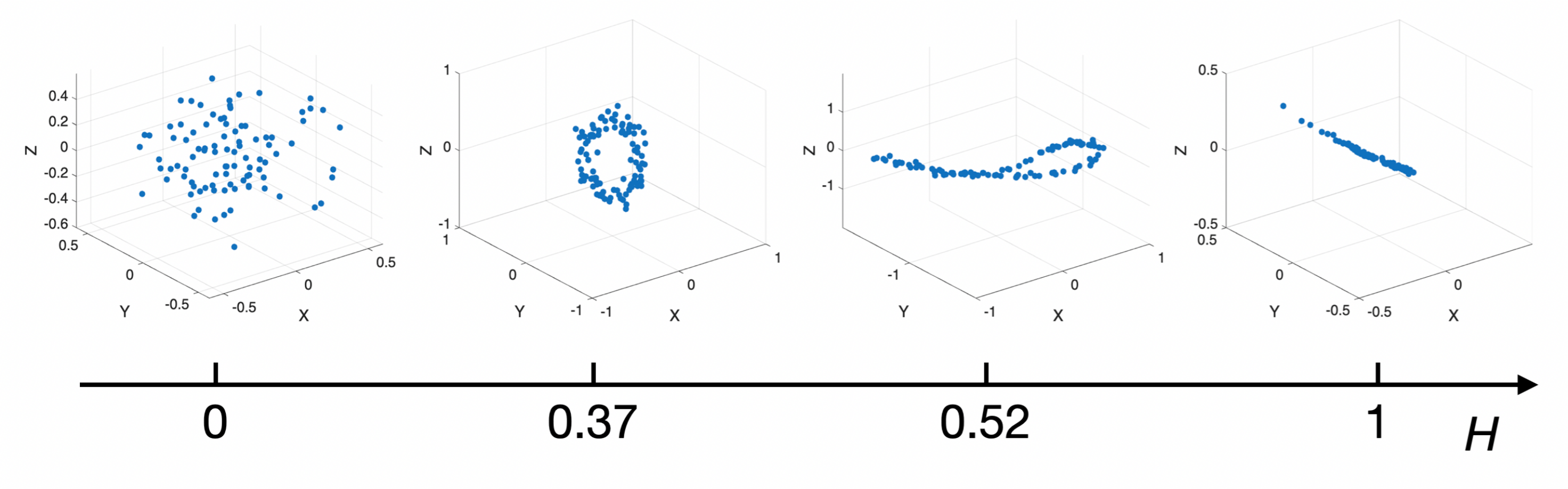}}
\caption{
Zoology of emergent collective patterns in $3$D as function of the non-reciprocity index ${H}$. Patterns correspond, from left to right, to a cloud ($\beta=\pi$), a ring ($\beta=2.06$), a polar, elongated, closed loop ($\beta=1.9$), and a worm ($\beta=1.5$). 
For all patterns $D=0.05$, except for closed polar loop where, 
to allow a better visualization of the pattern, $D=0.0072$. All other parameter values  as in Figs.~\ref{fig:1}-\ref{fig:5}.
 }
\label{fig:3D}
\end{center}
\end{figure*}

{\it Pattern metamorphosis--} 
We have already mentioned that, for a fixed parameter set, it is possible to see the emergence of different patterns depending on the initial conditions. 
Now, we explore whether we can go from one collective pattern to another by slowly varying the value of $\beta$. 
First, we observe that, starting with the same pattern, by repeating the same protocol where $\beta(t)$ changes from $\beta(t_1)=\beta_1$ to $\beta(t_f)=\beta_f$, we go through different patterns. 
The observed sequence of patterns is not fixed and depends on the realization of the noise between $t_1$ and $t_f$. 
For instance, starting with a worm, we can go to a cloud passing by a 3-Twist, or by a ring, or a 2-Twist and then a 3-Twist [see Fig.\ref{fig:5} (a)].  
On the other hand, we can start with a ring and reach a cloud by increasing the value of $\beta$ and, reversely, start with a cloud and get to a worm by passing through a ring that emerges at a different $\beta$ value. 
We can quantify these results estimating the probabilities of finding a configuration at the end of the time interval corresponding to a given $\beta$ value. It is worth noting that collective patterns survive the slow tuning of the field of view angle even outside the typical range in which these patterns appears, from random initial conditions, in the phase diagram of Fig. 1(b). 

From these numerical experiments, we learn that there is no reversibility and that there are strong hysteretic effects: if we start from a given pattern A at $\beta_0$ and, by tuning $\beta_0  \to \beta_1$, we reach pattern B at $\beta_1$, this does not imply that starting from pattern B and performing the reverse transformation $\beta_1 \to \beta_0$, we will end up with pattern A.\\

{\it Robustness of results and model extensions--} 
%
The diversity of complex patterns generated by non-reciprocal attraction highlights the relevance of such a simple rule as a pattern-forming mechanism for flocking. Importantly, this richness of flocking behaviors is not observed when particles interact through standard velocity-alignment mechanisms.
The minimal model explored here involves several idealizations (or simplifications): particles experience no repulsive forces, they are treated as point masses, and the analysis has so far been restricted to 2D.
Although we consider simplifications essential for gaining theoretical insight,  
in the following, we examine  to which extent the reported results depend on these specific model assumptions. 

First, we investigate the effect of including a short-range repulsive force. Provided that the repulsive core -- which can be associated with the particle size -- is sufficiently small, all patterns shown in Fig.~\ref{fig:2} remain stable.  Stronger repulsion leads to broader polar filaments and can cause the nematic ring to reorganize into two opposite polar filaments. Supplementary figures and details of the repulsive force implementation are provided in \cite{SuppInf}.
Note that, although we do not explicitly account for visual occlusion, the force $\mathbf{F}_i$ involves an average over the $n_i$ neighbors of the particle $i$. Consequently, the dynamics of the particle $i$ is identical whether it has a single neighbor or multiple neighbors aligned behind one another. Furthermore, visual occlusion becomes relevant only when particles come into very close proximity,  a situation that can generally be avoided by introducing an effective repulsion that keeps particles separated. 

Second, we examine the impact of the spatial dimension on the system dynamics. Using Eq.~(\ref{eq:3d})
we investigate the emergent collective behavior in 3D. Details of the implementation are provided in \cite{SuppInf}.
As the field of view $\beta$ is decreased, 
we observe cloud-like patterns for reciprocal interactions ($\beta = \pi$), nematic ring patterns for $\beta \sim 2.2$, closed elongated (locally) polar loops for $\beta \sim 1.9$, and worm-like patterns when $\beta < 1.5$ -- see Fig.~\ref{fig:3D}.
Interestingly, the 2-twist and 3-twist patterns observed in 2D appear to be replaced by elongated polar loops, suggesting the absence of singular topological points in three dimensions.
Even if a more detailed study of the 3D structures is required, 
the comparison of Figs.~\ref{fig:2} and~\ref{fig:3D} suggests that collective patterns  
in 2D and 3D are qualitatively similar.

{\it Conclusions--} 
Here, we have studied active particles with long-range interactions restricted only by a field of view. 
In sharp contrast to active systems with short-range interaction, the system does not possess 
a gas phase. 
Importantly, we have shown that the combined effect of attraction and a field of view leads particles 
to self-organize within a manifold of reduced dynamical dimensions, i.e., a complex collective pattern in dimensions 2 and 3. 
Furthermore, we have revealed the existence of a fundamental interplay between the topology and the transport properties of the emergent structures, whose morphology is regulated by the degree of non-reciprocity.
The obtained results shed light on the physics of single-species, non-reciprocal active particles and may help elucidate emergent, complex collective behaviors observed in animal groups and swarm robotics. 
An analytical understanding of the complex patterns that emerge in a system of active particles with non-reciprocal interactions -- beyond the numerical study performed here -- remains a major theoretical challenge for future research.
\\

E.B. was partially 
supported by CNPq (Grant No. 305008/2021-8 and 305984/2024-1) 
and FAPERJ (Grant No. 260003/005762/2024). 
F.P. acknowledges financial support from C.Y. Initiative of Excellence (grant Investissements d'Avenir ANR-16-IDEX- 0008), INEX 2021 Ambition Project CollInt, Labex MME-DII, projects 2021-258 and 2021-297, and projects ANR-22-CE30-0038 ``Push-pull" and (ANR-NSF) ANR-24-CE95-0002 ``MotDis". 

\bibliographystyle{apsrev}
\bibliography{biblio}


\end{document}